\documentclass[a4paper]{article}

\usepackage[english]{babel}
\usepackage[utf8x]{inputenc}
\usepackage[T1]{fontenc}
\usepackage{float}

\usepackage[a4paper,top=3cm,bottom=2cm,left=3cm,right=3cm,marginparwidth=1.75cm]{geometry}

\usepackage{amsmath}
\usepackage{algpseudocode}
\usepackage{graphicx}
\usepackage[colorinlistoftodos]{todonotes}
\usepackage[colorlinks=true, allcolors=blue]{hyperref}

\newcommand*{\Comb}[2]{{}^{#1}C_{#2}}%

\title{CoHSI I: Detailed properties of the Canonical Distribution for Discrete Systems such as the Proteome}
\author{Les Hatton, Gregory Warr}

\begin{document}
\maketitle

\begin{abstract}
The CoHSI (Conservation of Hartley-Shannon Information) distribution is at the heart of a wide-class of discrete systems, defining the length distribution of their components amongst other global properties.  Discrete systems such as the known \textbf{proteome} where components are proteins, \textbf{computer software}, where components are functions and \textbf{texts} where components are books, are all known to fit this distribution accurately.  In this short paper, we explore its solution and its resulting properties and lay the foundation for a series of papers which will demonstrate amongst other things, why the average length of components is so highly conserved and why long components occur so frequently in these systems.  These properties are not amenable to local arguments such as natural selection in the case of the proteome or human volition in the case of computer software, and indeed turn out to be inevitable global properties of discrete systems devolving directly from CoHSI and shared by all.   We will illustrate this using examples from the Uniprot protein database as a prelude to subsequent studies.
\end{abstract}

\section{Statement of computational reproducibility}
Most scientists will be aware of the growing problem of computational irreproducibility in the software-consuming sciences, and so as with our previous papers in this area \cite{HatTSE14,HattonWarr2015,HattonWarr2017}, this paper is accompanied by a complete computational reproducibility suite including all software source code, data references and the various glue scripts necessary to reproduce each figure, table and statistical analysis and then regress local results against a gold standard embedded within the suite to help build confidence in the theory and results we are reporting.  This follows the methods broadly described by \cite{Ince2012} and exemplified in a tutorial and case study \cite{HattonWarr2016}.  These reproducibility suites are currently available at http://leshatton.org/ until a suitable public archive appears, where they may be transferred.

\section{Introduction}

This paper is one of a series based on the theory we developed in \cite{HatTSE14,HattonWarr2015,HattonWarr2017} which by merging Information Theory in a novel way as a constraint into a Statistical Mechanics framework, demonstrates that a wide class of discrete systems, when split into components, exhibits underlying properties which are identical.  For example, the length distribution of proteins and the length distribution of software components are functionally identical and derive from the same underlying cause, a conservation principle.  This principle is the Conservation of Hartley-Shannon Information \cite{Hartley1928,Shannon1948,Shannon1949,Cherry1963} which we will abbreviate throughout as \textbf{CoHSI}.  This principle is sufficient without any further assumptions other than the standard one that all microstates in the Statistical Mechanics framework are equally likely, to predict all the subtle features of the length distributions in these highly disparate systems.  Related work includes that of Frank \cite{Frank2009} who produces a simple and consistent informational framework of the common patterns of nature using a maximum-entropy formulation.  This allows him to demonstrate how patterns such as Poisson, Gaussian and power-law follow directly from simple constraints on information.  For example, Frank demonstrates that any aggregation of hidden small-scale processes that preserves information only about the geometric mean attracts to the power law pattern.  Where we differ is in starting with a model of a specific kind of discrete system which matches many discrete systems such as the proteome and following its Hartley-Shannon information content through a Statistical Mechanics framework leading directly to a description of its length distribution.  This length distribution asymptotes to a power-law but exhibits other complex behaviour which closely matches the observed properties of systems such as the proteome.   

We consider a discrete system as consisting of M components or boxes, indexed $i=1,..,M$, such that the $i^{th}$ component contains $t_{i}$ indivisible tokens chosen from $a_{i}$ unique choices, which we will call the alphabet of the $i^{th}$ component.  We also assume that these indivisible tokens are distinguishable by the order in which they appear.  

This is the \textit{heterogeneous} model described by \cite{HattonWarr2017} using colored beads as tokens.  We note that there is a complementary \textit{homogeneous} model with each box containing beads of only one color with no two boxes having the same color, but we will not consider it further here as its solution is both explicit and simple and manifests itself as a pure power-law; this model therefore serves as a proof of Zipf's law \cite{Zipf35}.  For a comprehensive discussion of power-laws in nature generally, see for example \cite{Newman2006}.

\begin{figure}[ht!]
\centering
\includegraphics[width=0.5\textwidth]{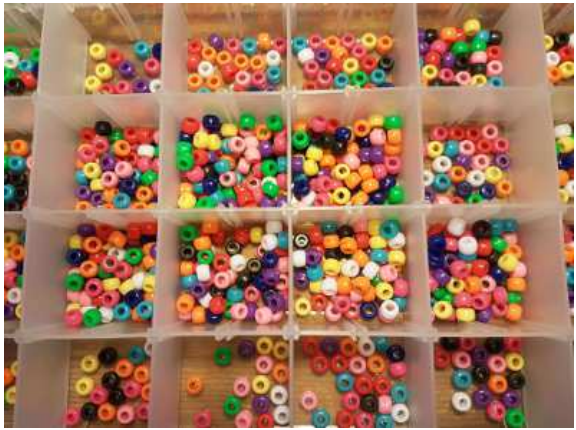}
\caption{\label{fig:beads}Exemplifying a heterogeneous system as described in the text.}
\end{figure}

The heterogeneous model can be simply exemplified by a box split up into $M$ compartments each of which contains colored beads as shown in Figure \ref{fig:beads}.  For an example from the physical world, each box might be a protein and each bead an amino acid chosen from the unique alphabet of amino acids \cite{HattonWarr2015} with the proviso that the order in which they appear is distinguishable, for example if they appear in sequence in their box to mirror the protein structure.

Using the standard definition of Hartley-Shannon Information as the log of the number of ways of arranging the beads in each box and introducing this as a constraint in a Statistical Mechanics argument \cite{HattonWarr2017}, then gives the overwhelmingly most likely distribution of lengths $t_{i}$ as being the implicit solution of

\begin{equation}
\log t_{i} = -\alpha -\beta ( \frac{d}{dt_{i}} \log N(t_{i}, a_{i}; a_{i} ) ),    \label{eq:minifst}
\end{equation}

We note in passing that this uses the simplest form of Stirling's approximation to $\log (t_{i} !) \approx t_{i} \log t_{i} - t_{i}$.
Since we are particularly interested in solutions for small $t_{i}$, we chose to use Ramanujan's form \cite{Ramanujan1988} and actually solved the following (although it makes little difference in practice)

\begin{equation}
\log t_{i} + \frac{1 + 8 t_{i} + 24t_{i}^{2}}{6(t_{i} + 4 t_{i}^{2} + 8 t_{i}^{3})} = -\alpha -\beta ( \frac{d}{dt_{i}} \log N(t_{i}, a_{i}; a_{i} ) ),    \label{eq:minif}
\end{equation}

Here $N(t_{i}, a_{i}; a_{i})$ is the number of different ways in which the $t_{i}$ beads in the $i^{th}$ box chosen from a unique alphabet of $a_{i}$ beads can be arranged where order is important.  It can also be visualised as the number of boxes of chocolates containing $t_{i}$ chocolates chosen with repetition from $a_{i}$ unique chocolates such that at least one of each of the $a_{i}$ is included and the order of chocolates in each box is significant \cite{HattonWarr2017}.

The constants $\alpha, \beta$ are the Lagrange undetermined multipliers from the Statistical Mechanics formulation.  We will study how the solution changes with these later.  In traditional formulations of Statistical Mechanics these decouple with $\alpha$ handling the normalisation and $\beta$ the shape.  For the CoHSI equation, this is only true for $t_{i} \gg a_{i}$ as we shall see.

To give an example of $N(t_{i}, a_{i}; a_{i})$, suppose we have a box of $t_{i} = 5$ beads such that it contains \textit{exactly} $a_{i} = 2$ different beads of colors A and B.  The total number of ways this box can be arranged is then

\begin{equation}
 N(5,2;2) = \frac{5!}{1!4!} + \frac{5!}{4!1!} + \frac{5!}{3!2!} + \frac{5!}{2!3!} = 30			\label{eq:chocs52}
\end{equation} 

We have used the $; a_{i}$ nomenclature because it turns out that this function depends on the additive compositions of numbers (the number 5 can be decomposed into additive compositions of size 2 in 4 ways: 1+4=5, 4+1=5, 2+3=5, 3+2=5).  For example, the first term on the rhs of (\ref{eq:chocs52}) is the number of ways a box of 5 beads can be constructed using 1 of color A and 4 of color B, re-iterating that order is significant.  $N(t_{i}, a_{i}; a_{i})$ can then be evaluated recursively by the following relation \cite{HattonWarr2017}

\paragraph{}

\begin{minipage}{\textwidth}
\begin{algorithmic}
\For {$t_{i} = 1,..,t_{i} (MAX)$}
  \For {$a_{i} = 1,..,t_{i}$}
    \State $N(t_{i}, 1; 1) = 1;$
    \For {$i = 1,..,(a_{i} - 1)$}
      \State $N(t_{i}, a_{i}; i) \gets \Comb{a_{i}}{i} N(t_{i}, i; i)$
    \EndFor
    \State $N(t_{i}, a_{i}; a_{i}) \gets a_{i}^{t_{i}} - \sum_{i=1}^{a_{i}-1} N(t_{i},a_{i}; i)$
  \EndFor
\EndFor
\end{algorithmic}
\end{minipage}

\paragraph{}

We give an indication of how this recursion works by showing a few examples in Table \ref{tab:chocs}.  The earlier example for a box of 5 beads made from 2 unique types of bead is shown as N(5,2;2) = 30, as expected.

\begin{table}[h!]
\centering
\begin{tabular}{p{2cm}p{8cm}p{2cm}}
\hline
N($t_{i}, a_{i}; a'_{i}$) & Formula & Different possible boxes \\
\hline
N(5,1;1) & 1 & 1 \\
\hline
N(5,2;1) & $\Comb{2}{1}$ N(5,1;1) & 2 \\
N(5,2;2) & $2^{5}$ - N(5,2;1) & 30 \\
\hline
N(5,3;1) & $\Comb{3}{1}$ N(5,1;1) & 3 \\
N(5,3;2) & $\Comb{3}{2}$ N(5,2;2) & 90 \\
N(5,3;3) & $3^{5}$ - N(5,3;1) - N(5,3;2) & 150 \\
\hline
N(5,4;1) & $\Comb{4}{1}$ N(5,1;1) & 4 \\
N(5,4;2) & $\Comb{4}{2}$ N(5,2;2) & 180 \\
N(5,4;3) & $\Comb{4}{3}$ N(5,3;3) & 600 \\
N(5,4;4) & $4^{5}$ - N(5,4;1) - N(5,4;2) - N(5,4;3) & 240 \\
\hline
N(5,5;1) & $\Comb{5}{1}$ N(5,1;1) & 5 \\
N(5,5;2) & $\Comb{5}{2}$ N(5,2;2) & 300 \\
N(5,5;3) & $\Comb{5}{3}$ N(5,3;3) & 1500 \\
N(5,5;4) & $\Comb{5}{4}$ N(5,4;4) & 1200 \\
N(5,5;5) & $5^{5}$ - N(5,4;1) - N(5,4;2) - N(5,4;3) - N(5,4;4) & 120 \\
\hline
\end{tabular}
\caption{Illustrating how the recursion works for calculating the number of ways a box of beads can be constructed given the total number of beads and the total number of uniquely colored beads where bead order matters.}
\label{tab:chocs}
\end{table}

As was pointed out in \cite{HattonWarr2017}, (\ref{eq:minif}) tends asymptotically to the following when $t_{i} \gg a_{i}$.

\begin{equation}
\log t_{i} = -\alpha -\beta ( \log a_{i} )    \label{eq:minip}
\end{equation}

Equation (\ref{eq:minip}) can then be integrated to give

\begin{equation}
t_{i} = e^{-\alpha} a_{i}^{-\beta}    \label{eq:minipint}
\end{equation}

This corresponds to an explicit pdf as it is normalisable for $\beta > 1$ over an infinite support and generally over a finite support, and is non-negative everywhere.  To see this, we note at this point that since the total number of beads $T$ is constrained in this formulation \cite{HattonWarr2017}, that

\begin{equation}
\sum_{i=1}^{M} t_{i} = T
\end{equation}

It is clear then from (\ref{eq:minipint}) that $\alpha$ is associated with normalisation so that it becomes a pdf and that $\beta$ is a parameter associated with the shape of the distribution, which is of course a power-law.  \textit{Of considerable importance is that $T$ does not appear in either the implicit pdf (\ref{eq:minif}) or its asymptotic explicit form (\ref{eq:minip}).  In other words, the solutions are scale-independent.}

So we know that the implicit pdf solution of (\ref{eq:minif}) will tend to the power-law (\ref{eq:minip}) when $t_{i} \gg a_{i}$.  It turns out by simple algebraic manipulation \cite{HattonWarr2015} that the power-law solution (\ref{eq:minip}) has a dual relation 

\begin{equation}
a_{i} \sim t_{i}^{-1/\beta}    \label{eq:minipina}
\end{equation}

(\ref{eq:minipint}) and (\ref{eq:minipina}) imply that \textit{both} the unique alphabet and the length distributions follow a power-law asymptotically.  This is explained in more detail and then demonstrated emphatically on the Uniprot database in \cite{HattonWarr2015}.  For our purposes, this means that we can generate either from the full CoHSI equation (\ref{eq:minifst}) or (\ref{eq:minif}).

\begin{figure}[ht!]
\centering
\includegraphics[width=0.5\textwidth]{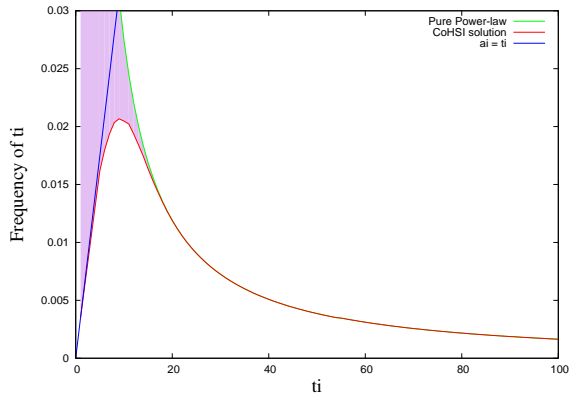}
\caption{\label{fig:fullandpower}Illustrating how the implicit solution of the CoHSI length distribution asymptotes to a power-law.}
\end{figure}

Here we focus on the CoHSI length distribution and illustrate this by numerically integrating (\ref{eq:minif}) by specifying a range of $t_{i}$ and calculating the corresponding $a_{i}$ and then normalising to show how the CoHSI solution produces a pdf with the characteristic sharp unimodal peak before asymptoting into the power-law (\ref{eq:minip}).  This is shown in Figure \ref{fig:fullandpower}.  The shape is reminiscent of both gamma and lognormal distributions but has the precise power-law long tail.

The shaded zone of Figure \ref{fig:fullandpower} shows the difference between the power-law solution (in green) and the full CoHSI solution (in red).  For boxes where $t_{i} \gg a_{i}$, the two are indistinguishable as we expect because the number of ways the box can be arranged remains at $a_{i}^{t_{i}}$ as $t_{i}$ continues to increase as there is little risk in missing out any of the $a_{i}$ because the number of choices $t_{i}$ is large.  We re-iterate that the information model assumes that each of the $a_{i}$ must always be used at least once by definition whatever the value of $t_{i}$.  The $a_{i}^{t_{i}}$ term is of course the engine of the naturally emergent power-law.

However as $t_{i}$ decreases, something very interesting happens.  The inclusion of the $d/dt_{i}$ gradient of the $\log N(t_{i}, a_{i}; a_{i})$ function causes the solution to depart increasingly from the power-law.  Physically the reason for this is that the number of ways of arranging the beads in the boxes \textit{guaranteeing that their unique alphabet remains as $a_{i}$} is greatly reduced and this pulls the solution pdf up into the sharp peak and then very rapidly down to zero.  (Note that $t_{i}$ cannot be smaller than $a_{i}$ in (\ref{eq:minif}) because no box can contain fewer beads than its unique alphabet.) 

\begin{figure}[ht!]
\centering
\includegraphics[width=0.5\textwidth]{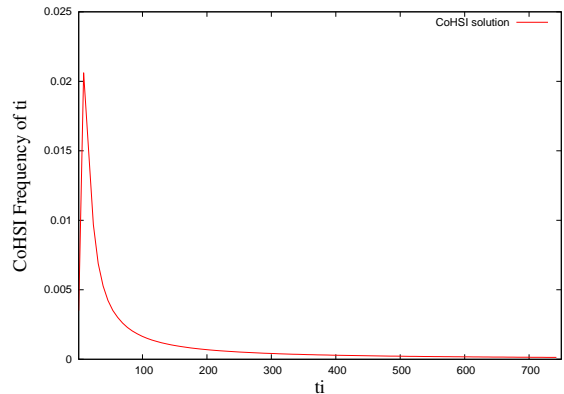}
\caption{\label{fig:singlefar}Illustrating a typical solution of the CoHSI equation (\ref{eq:minif}) for $\alpha = 4, \beta = 0.8$.}
\end{figure}

Before we consider the properties in detail, we also show a typical numerical solution of (\ref{eq:minif}) as Figure \ref{fig:singlefar} without other encumbrances.  Here $\alpha = 4, \beta = 0.8$.

\begin{figure}[ht!]
\centering
\includegraphics[width=0.5\textwidth]{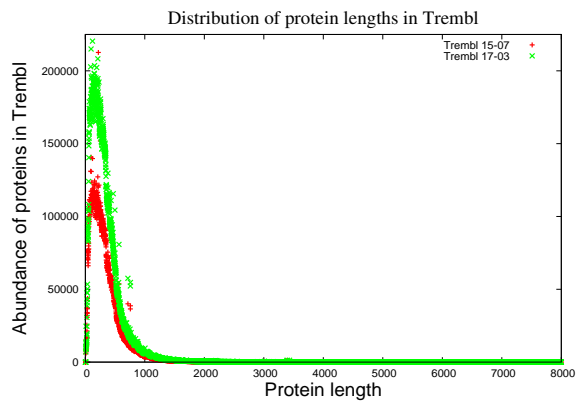}
\caption{\label{fig:cdata}Illustrating the length distribution of the proteins in TrEMBL versions 15-07 and 17-03}
\end{figure}

To see the close agreement with length distributions in real data, Figure \ref{fig:cdata} shows the length distributions of the proteins in two versions of the full TrEMBL database\footnote{https://uniprot.org/}, separated by 18 months in which time the database doubled in size.  Both distributions asymptote to an extraordinarily accurate power-law as predicted by the scale-independent (\ref{eq:minif}) and (\ref{eq:minip}) and this and many other examples are shown in \cite{HattonWarr2017}.

The rest of this paper will explore the solution and the properties of (\ref{eq:minif}) as the undetermined Lagrange parameters $\alpha, \beta$ are varied.

\section{Solving the CoHSI equation}
It should be re-iterated first that (\ref{eq:minif}) is an implicitly defined pdf which asymptotes to an explicit pdf for boxes which are large compared with their unique alphabet.  Implicitly defined pdfs are unusual but not unique as Tsallis entropy \cite{Tsallis1988,Tsallis1999}, can also lead to an implicitly defined pdf.  The implicit nature leads to a few challenges in its solution.

Because of the recursively defined nature of $N(t_{i}, a_{i}; a_{i})$ and the fact that the function is calculable only at integer values of $t_{i}, a_{i}$, initially the log of this function was calculated over a grid at integer points, $t_{i} = 1,..,100; a_{i} = 1,..,50$.  Anticipating the need for real values as part of a repeated bisection method for solving the implicit equation (\ref{eq:minif}), the derivative $d/dt_{i}$ was computed numerically on this grid and interpolated between when necessary.

The bisection method converges rapidly as can be seen in Figure \ref{fig:fullandpower} until $t_{i}$ is around 4 at which point the solution curve gets sufficiently close to the boundary defined by $t_{i} = a_{i}$, that the bisection method fails due to discretisation errors in the computation of the derivative, (second order differences are used everywhere except the boundaries where first-order forward differences are used).  This could probably be resolved but is unnecessary as the solution is perfectly adequate over the range of $t_{i} \ge 4$.

\section{Exploring the CoHSI equation}
The two disposable parameters are $\alpha, \beta$.  As we saw above, in the asymptotic power-law form, $\alpha$ is a pure normalising parameter allowing the curve to be normalised such that the area under it is 1 as required for a pdf.  The parameter $\beta$ determines the shape of the power-law.  In essence, they are completely decoupled - once a choice of $\beta$ is made, only one $\alpha$ is available, the one which correctly normalises the pdf to have an integral of 1 over its support or range.

However as $t_{i}$ decreases, things become more complicated.  The explicit power-law pdf (\ref{eq:minip}) mutates into the full CoHSI equation (\ref{eq:minif}) and \textbf{both} $\alpha$ and $\beta$ contribute to normalisation and shape.  In other words for either fixed $\alpha$ \textit{or} fixed $\beta$, a range of values of the other is available giving a much richer structure for these smaller $t_{i}$.

\begin{figure}[ht!]
\centering
\includegraphics[width=0.5\textwidth]{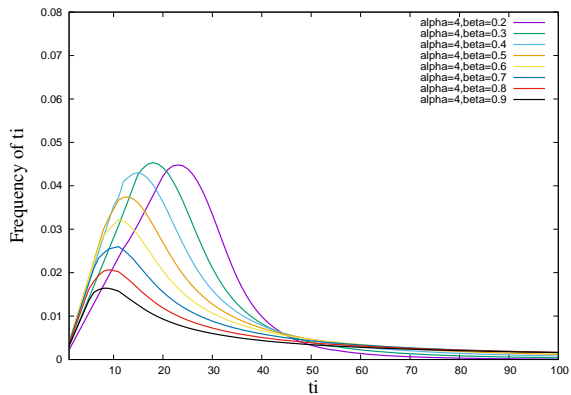}
\caption{\label{fig:fixedalpha}Illustrating solutions of the CoHSI equation (\ref{eq:minif}) for $\alpha = 4, \beta = 0.2,0.3,..,0.9$.}
\end{figure}

\begin{figure}[ht!]
\centering
\includegraphics[width=0.5\textwidth]{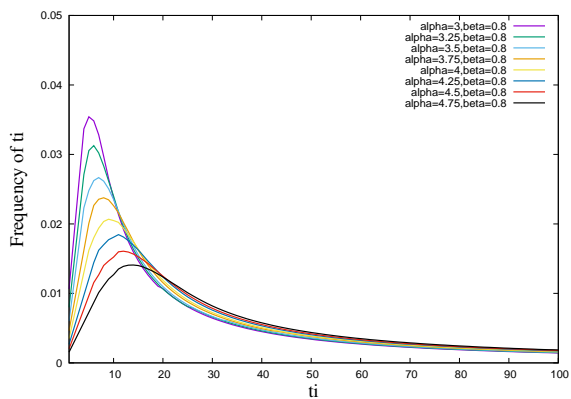}
\caption{\label{fig:fixedbeta}Illustrating solutions of the CoHSI equation (\ref{eq:minif}) for $\alpha = 3,3.25,..,4.75, \beta = 0.8$.}
\end{figure}

To see this, Figures \ref{fig:fixedalpha} and \ref{fig:fixedbeta} show the CoHSI solution for (fixed $\alpha$, variable $\beta$) and (fixed $\beta$, variable $\alpha$) respectively.  Each one is a correctly normalised pdf.

In Figure \ref{fig:fixedalpha}, we see that for fixed $\alpha$ and variable $\beta$, the peak moves to the left with increasing $\beta$, whilst the power-law ($t_{i} \geq 40$) tends to zero more slowly.

In Figure \ref{fig:fixedbeta}, we see that for fixed $\beta$ and variable $\alpha$, the peak this time moves to the right with increasing $\alpha$, whilst the power-law ($t_{i} \geq 40$) is largely unaffected as we would expect.

This coupling means that the parameters $\alpha,\beta$ no longer have an intuitively obvious connection with the length distributions.  However, the slope of the power-law as derived from the ccdf throughout this work \textit{does} have an intuitive connection in that it can be extracted directly from real data.  In order to relate it to the parameters $\alpha,\beta$, we must solve the CoHSI equation for a grid of values of $\alpha,\beta$ and for each, we compute the ccdf and extract its power-law slope.  This we do now.

\subsection{$\alpha$, $\beta$ and the power-law slope}
\begin{figure}[H]
\centering
\includegraphics[width=0.5\textwidth]{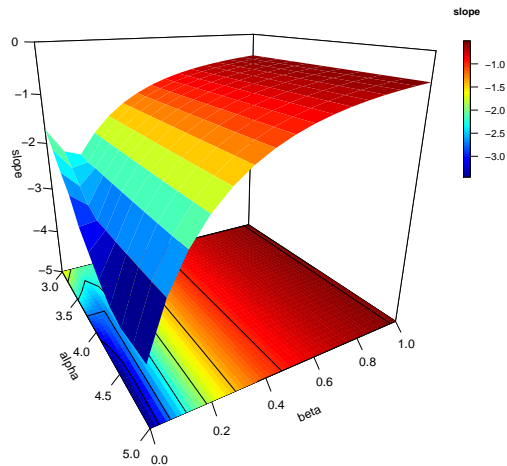}
\caption{\label{fig:slopealphabeta}A perspective plot of the actual slope of the power-law derived from a R lm() analysis of the ccdf of each length distribution for a grid of values of $\alpha, \beta$ over a 2-decade range $t_{i} = 20,..,2000$.}
\end{figure}

\begin{figure}[H]
\centering
\includegraphics[width=0.5\textwidth]{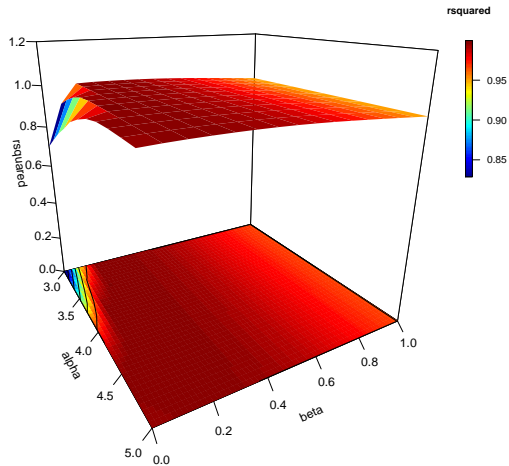}
\caption{\label{fig:rsquaredalphabeta}A perspective plot of the adjusted $R^{2}$ fit of the power-law slope of the power-law derived from a R lm() analysis of the ccdf of each length distribution for a grid of values of $\alpha, \beta$ over a 2-decade range $t_{i} = 20,..,2000$.  As can be seen, this is very close to 1 over most of the range indicating a high-quality power-law fit.}
\end{figure}

Asymptotically the slope of the $t_{i}$ pdf corresponds approximately to $1/\beta$ \cite{HattonWarr2015}, however this depends on the unique alphabet used.  Here we explore the actual slopes found in real data to see how they compare with the values of the Lagrangian undetermined parameters $\alpha,\beta$ which appear in the CoHSI equation. The ccdf power-law slope of the complete TrEMBL dataset in that paper is $-3.13$.  Figure \ref{fig:slopealphabeta} was calculated by constructing the ccdf of the length distribution for a grid of values of $\alpha, \beta$.  The goodness of fit in terms of adjusted $R^{2}$ is shown as Figure \ref{fig:rsquaredalphabeta}.  The closer this is to 1, the better the power-law approximation.

As can be seen, the slope value observed in TrEMBL corresponds to large $\alpha$ and small $\beta$ and the fit in this area has a value of adjusted $R^{2}$ close to 1.

%
%
\subsection{Averages}
Defining the average value of a distribution such as that shown in Figure \ref{fig:singlefar} is problematic as the distribution is significantly right-skewed with a sharp unimodal peak followed by a long power-law tail.  Three common methods of measuring the average or "centre" of the distribution are the mean, the median and the mode.  For symmetric unimodal pdfs, these three measures are coincident, but when the distribution is asymmetric, they are not.

When we think of an average, we are probably thinking of the most likely value to encounter.  When faced with the classic bell-shaped normal distribution, the most likely value is the mean, which happens to be coincident with both the median and the mode as mentioned above.  This expectation becomes distinctly vague when faced with distributions as skewed as that of the CoHSI distribution Figure \ref{fig:singlefar}.  It is certainly possible to apply standard statistical inference methods to such distributions using for example a logarithmic scaling in $t_{i}$ to approximate a lognormal distribution\footnote{https://www.ma.utexas.edu/users/mks/statmistakes/skeweddistributions.html}.  Here we simply explore graphically how the three common measures, mean, median and mode all behave over a typical grid of values of $\alpha,\beta$.  These can then be related simply to the actual ccdf slope using Fig. \ref{fig:slopealphabeta}.

\subsubsection{The mean}
The mean is often used as a measure of the "Centre" of a distribution. As noted by \cite{XuJune2006}, this is highly conserved within both Prokaryotes and Eukaryotes but is divergent across these two domains of life.  We might in fact expect that distributions such as Figure \ref{fig:singlefar} would have a mean value which did not vary much if the overall shape is relatively conserved.

\begin{figure}[ht!]
\centering
\includegraphics[width=0.5\textwidth]{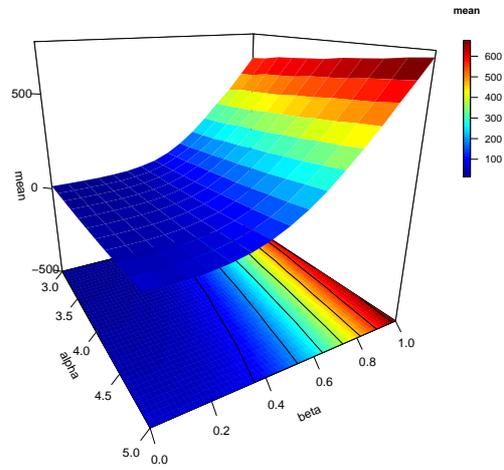}
\caption{\label{fig:meanalphabeta}A perspective plot of the variation of the mean with alpha and beta.}
\end{figure}

Figure \ref{fig:meanalphabeta} shows the behavior of the mean of the CoHSI distribution for a range of values of $\alpha,\beta$.  We note that for values of $\beta$ which vary within a small range as is common in real systems, the mean does not vary much with $\alpha$ as expected.

\subsubsection{The median}

\begin{figure}[ht!]
\centering
\includegraphics[width=0.5\textwidth]{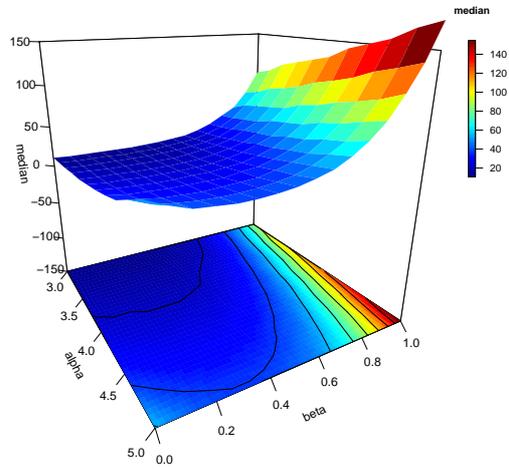}
\caption{\label{fig:medianalphabeta}A perspective plot of the variation of the mean with alpha and beta.}
\end{figure}

Over the same range of $\alpha,\beta$, Figure \ref{fig:medianalphabeta} demonstrates that as expected for a skewed distribution, the median is rather less sensitive but follows the same overall pattern as the mean, increasing towards the high values of both $\alpha,\beta$.

\subsubsection{The mode}

\begin{figure}[H]
\centering
\includegraphics[width=0.5\textwidth]{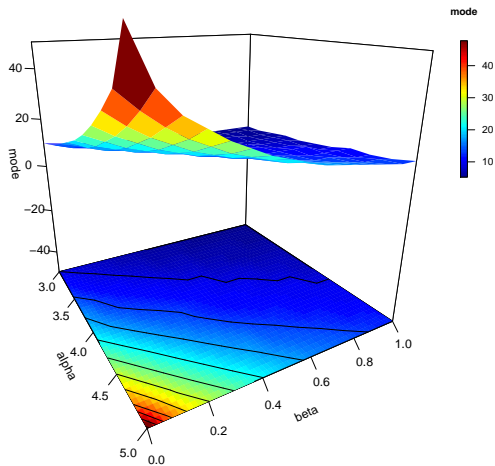}
\caption{\label{fig:modealphabeta}A perspective plot of the variation of the mean with alpha and beta.}
\end{figure}

In contrast, the mode, which represents the peak value and is therefore perhaps most eye-catching to the viewer behaves differently as can be seen in Figure \ref{fig:modealphabeta}.  Here the mode increases towards a peak for high $\alpha$ but low $\beta$.  In paper II of this series we will extend this discussion over the full range of proteins in TrEMBL.

\subsection{Skewness and long components}
We simply comment here rather than discuss as this will be taken up in detail in paper III in this series.  The right-skewed nature of the CoHSI distribution implies inevitably, that long components will occur surprisingly often for those used to exponential distributions such as the Gaussian, (``events that are effectively 'impossible' (negligible probability under an exponential distribution) become practically commonplace under a power-law distribution.'' \cite{Clauset2011}).   This property has been widely observed in both \cite{Louridas:2008:PLS:1391984.1391986,HattonWarr2017}, but the startling implication here is that long components are a global property due to the properties of the CoHSI equation and have little if anything to do with natural selection in the case of proteins or human volition in the case of software components. 

We will return to these diagrams in later papers as we home in on properties such as observed average length across the domains of life.

%
%
\section{Conclusions}
The CoHSI equation described in the text (\ref{eq:minif})

\[
\log t_{i} + \frac{1 + 8 t_{i} + 24t_{i}^{2}}{6(t_{i} + 4 t_{i}^{2} + 8 t_{i}^{3})} = -\alpha -\beta ( \frac{d}{dt_{i}} \log N(t_{i}, a_{i}; a_{i} ) ),
\]

successfully models the length distribution of a wide class of discrete systems including the proteome, computer software and texts \cite{HattonWarr2017}.  Here we have explored its solution and its properties in some detail so that these may form the basis of further studies applying these principles to various discrete systems, notably the proteome.

In addition to its existing properties,

\begin{itemize}
\item The CoHSI equation embodies an implicitly defined pdf which accurately models the length distribution of proteins and other discrete systems.
\item The CoHSI equation naturally and inevitably asymptotes to an explicit power-law pdf for large components in any of these systems as is observed.
\item The CoHSI equation is parameterised by 2 parameters $\alpha,\beta$ devolving from the Statistical Mechanics argument underlying it.
\end{itemize}

\cite{HattonWarr2017}, we have shown here that

\begin{itemize}
\item $\alpha,\beta$ \textit{decouple} for large components with $\alpha$ controlling normalisation and $\beta$ the power-law slope.
\item $\alpha,\beta$ are \textit{coupled} for smaller components with both contributing to the shape and position of the unimodal peak giving a rich set of behaviors.
\item The mean and the median behave similarly with both increasing as $\alpha,\beta$ increase.
\item The mode behaves in a retrograde manner increasing for increasing $\alpha$ and decreasing $\beta$.
\item The CoHSI equation is scale-independent as there is no explicit dependence on $T$.
\item The actual power-law slope of the ccdf bears a complex relationship with the Lagrangian parameters $\alpha, \beta$ with typical real values in TrEMBL corresponding to values of $\alpha \sim 5$ and $\beta \sim 0.1$.
\end{itemize}

We re-iterate that the only assumptions made are that all microstates in the sense of Statistical Mechanics are equally likely, and that there are a reasonable number of components, $M$.

In following papers, we will show how these properties can explain observed but previously unexplained properties of the proteome and other discrete systems in spite of the parsimony of the underlying information model.

\section{Acknowledgements}
The 3D graphics were produced using the plot3D package of R due to Karline Soetaert.


%
%



\bibliographystyle{alpha}
\bibliography{bibliography}

\end{document}